\documentclass[a4paper,preprintnumbers]{jpconf}
\usepackage{graphicx}
\usepackage{amsmath,amssymb}
\usepackage[square,sort&compress,numbers]{natbib}

\newcommand{\blockgen}{BlockGen}
\newcommand{\blockgenco}{\blockgen-CO}
\newcommand{\blockgencosummed}{\blockgenco$_\Sigma$}

\newcommand{\blockgencdsampled}{\blockgen-CD$_\text{MC}$}

\begin{document}
\begin{flushright}
  FERMILAB-CONF-22-071-T
\end{flushright}

\title{Generators and the (Accelerated) Future}

\author{J Isaacson}

\address{Particle Theory Department, Fermi National Accelerator Laboratory, Batavia, IL 60510, USA}

\ead{isaacson@fnal.gov}

\begin{abstract}
With the High Luminosity LHC coming online in the near future, event generators will need to provide very large event samples to match the experimental precision. Currently, the estimated cost to generate these events exceeds the computing budget of the LHC experiments. To address these issues, the computing efficiency of event generators need to be improved. Many different approaches are being taken to achieve this goal. I will cover the ongoing work on implementing event generators on the GPUs, machine learning the matrix element, machine learning the phase space, and minimizing the number of negative weight events.
\end{abstract}

\section{Introduction}
Collider experiments are developed by the high energy physics community to understand the universe at the smallest distances.
Event generators are tools that have been developed to connect the theory calculations to the experimental measurements.
This connection is made through simulating the collisions event-by-event through the use of Monte-Carlo methods.
The event generators can be broken down into separate components describing the physics at different scales involved in the collision~\cite{Webber:1986mc,Buckley:2011ms}.
Starting from the highest scale and working to the lowest scale, these consist of the hard matrix element, parton showers, and hadronization.
Our fundamental theoretical understanding is encoded in the hard matrix element through matrix element generators.
Matrix element generators are able to compute processes at tree-level~\cite{Kanaki:2000ey,Papadopoulos:2000tt,Krauss:2001iv,Gleisberg:2008fv,Maltoni:2002qb,Alwall:2007st,Alwall:2011uj,Mangano:2002ea}, one-loop level~\cite{Ossola:2007ax,Gleisberg:2007md,Berger:2008sj,Bevilacqua:2011xh,Cascioli:2011va,Cullen:2014yla,Alwall:2014hca,Actis:2016mpe}, and can handle a broad range of Beyond the Standard Model scenarios in an automated fashion~\cite{Christensen:2009jx,Degrande:2011ua,Staub:2013tta}.

\begin{figure}[h]
    \includegraphics[width=0.48\textwidth]{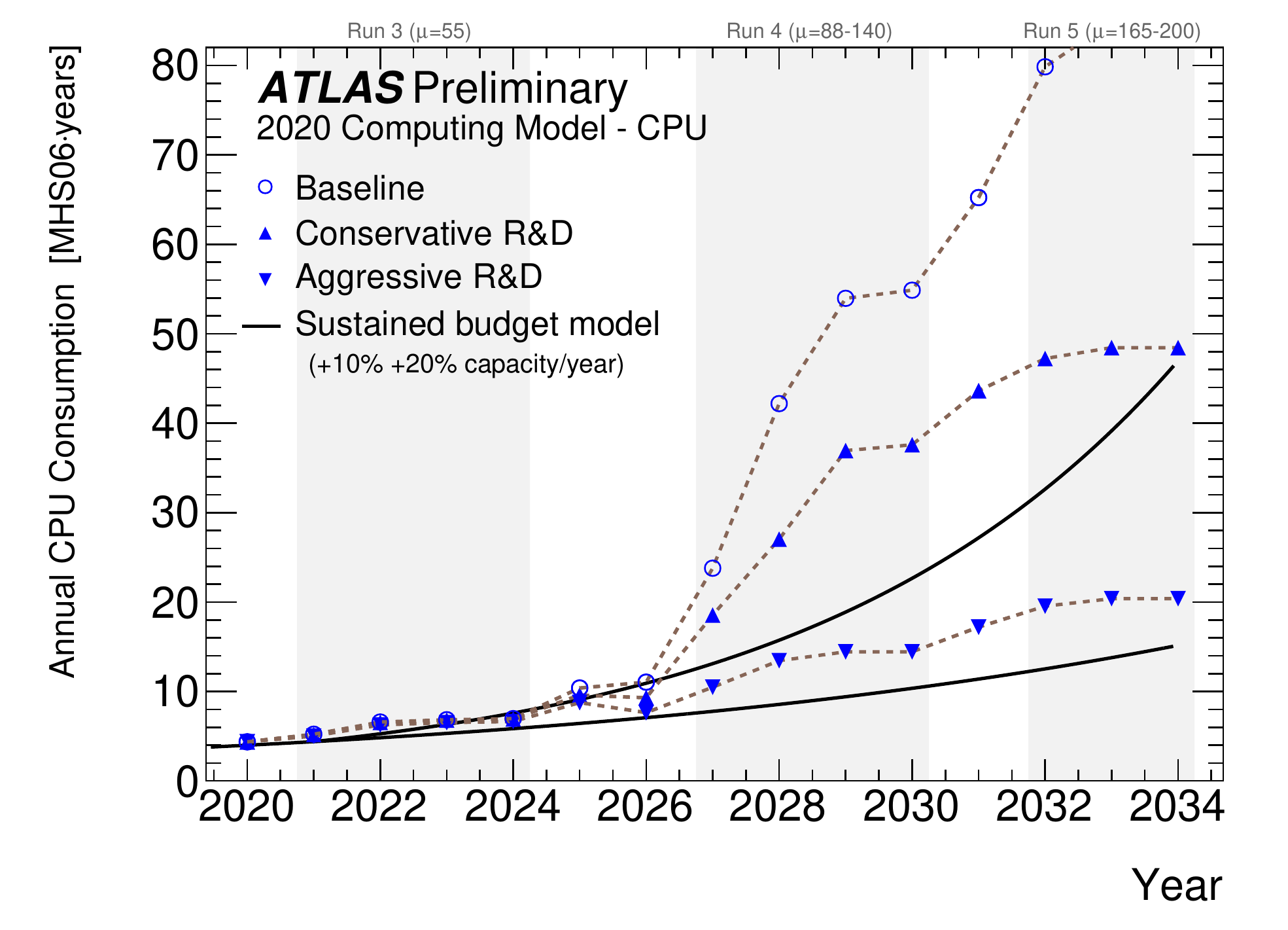}
    \includegraphics[width=0.47\textwidth]{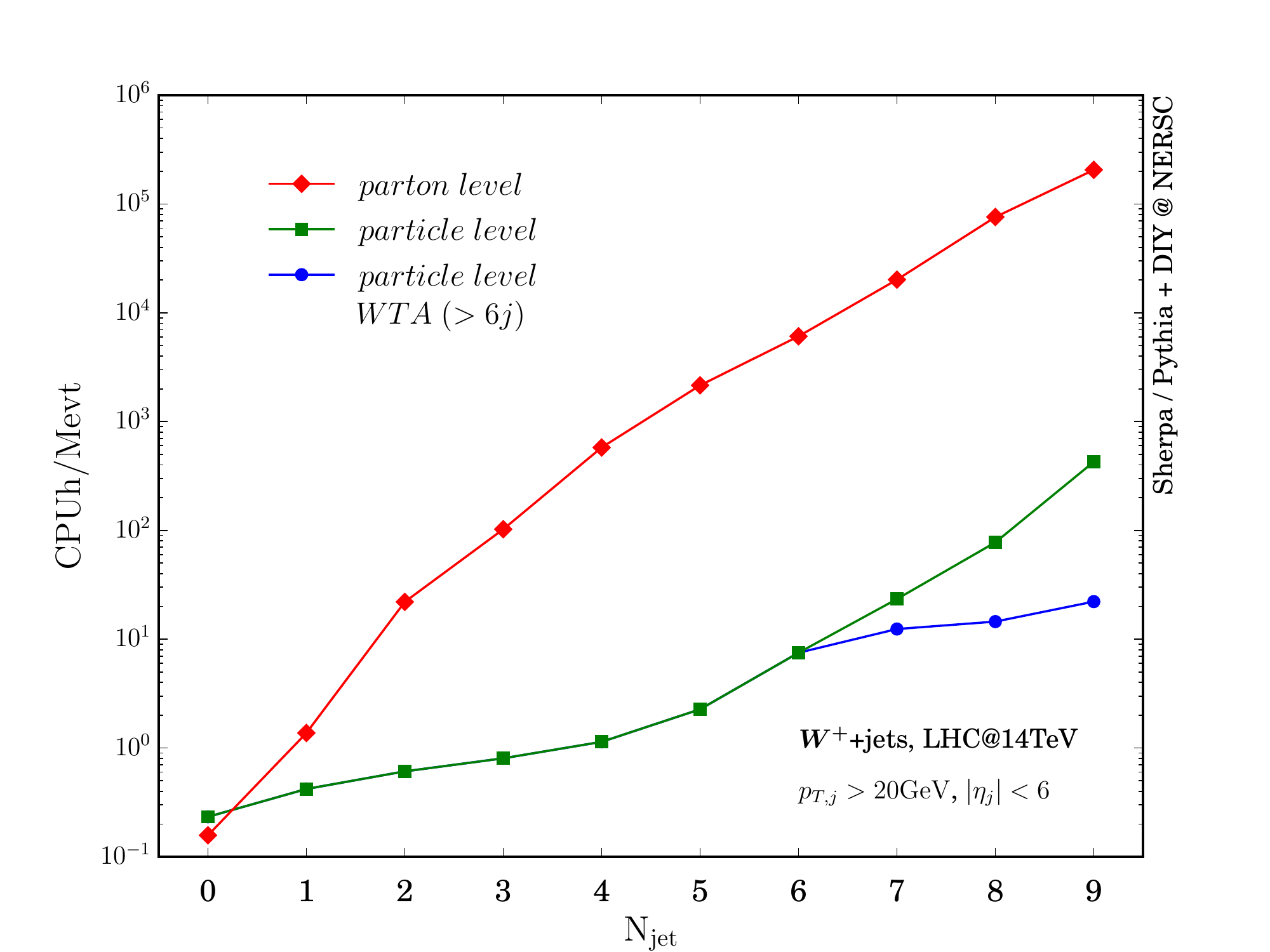}
    \caption{Left: Estimated computing requirements for current (open circles), conservative speedups (closed vertical triangles), and aggressive speedups (closed inverted triangles) as a function of time. Additionally, estimates of the expected computing budget are given as solid black lines. The figure is reproduced from Ref.~\cite{Calafiura:2729668}.
    Right: CPU hours required to generate one million events as a function of number of final state jets. The calculation of the hard matrix element is shown in ref, and the calculation of the parton shower is shown in green and blue. This figure is reproduced from Ref.~\cite{Hoche:2019flt}.}
    \label{fig:cpu_budget}
\end{figure}

As typically is the case, increasing the flexibility of these programs comes at a cost of computational efficiency. 
The alternative to these automated tools is to obtain analytic calculations for each process. While this is feasible for low multiplicity, there are no analytic results for high multiplicity processes.
Therefore, the trade-off in efficiency is acceptable in order to simulate the needed higher multiplicity events.
High multiplicity events will become even more important with the high-luminosity phase of the Large Hadron Collider (LHC) starting up in the near future. 
The datasets collected in this phase of the LHC will have an unprecedented number of events.
Ensuring that the event generator statistics are not the dominant uncertainty in these analyses requires a similar sized dataset. 
However with current techniques, this would far exceed the computational budget for the experiments (see the left panel of Fig.~\ref{fig:cpu_budget}). The cost of these events are dominated by the hard matrix element calculation, which scales exponentially as a function of the multiplicity (see the right panel of Fig.~\ref{fig:cpu_budget}).

To address these computing limitations, many different solutions are being investigated.
Here we will focus on four major categories of advancements being studied.
The first is developing matrix element generators that can take advantage of hardware accelerators such as graphics processing units (GPUs), and will be discussed in Sec.~\ref{sec:GPU}.
The second is using machine learning to develop a quick but accurate estimate of the matrix element to reduce the need to perform the full calculation. Details of this approach can be found in Sec.~\ref{sec:ME_ML}.
Thirdly, machine learning can also be used to more efficiently sample the phase space integrals involved in simulating the events for the colliders, which is explained further in Sec.~\ref{sec:PS_ML}.
Finally, when going beyond tree-level descriptions, negative weights are unavoidable. Negative weights drastically decrease the effective sample size. Methods to reduce the negative weight fraction are reviewed in Sec.~\ref{sec:negative}.

\section{GPU Matrix Element Generators}\label{sec:GPU}

There are three major efforts investigating the use of GPUs for matrix element calculations.
Each group takes a different approach in handling the conversion from CPU codes to GPU codes. These approaches are the ones developed in \blockgen~\cite{Bothmann:2021nch}, MadFlow~\cite{Carrazza:2021gpx}, and MadGraph-GPU~\cite{Valassi:2021ljk}.

\subsection{\blockgen}

\begin{figure}[t]
    \centering
    \begin{minipage}[b]{0.495\textwidth}
    \includegraphics[width=\linewidth,clip,trim=0mm 45mm 0mm 35mm]{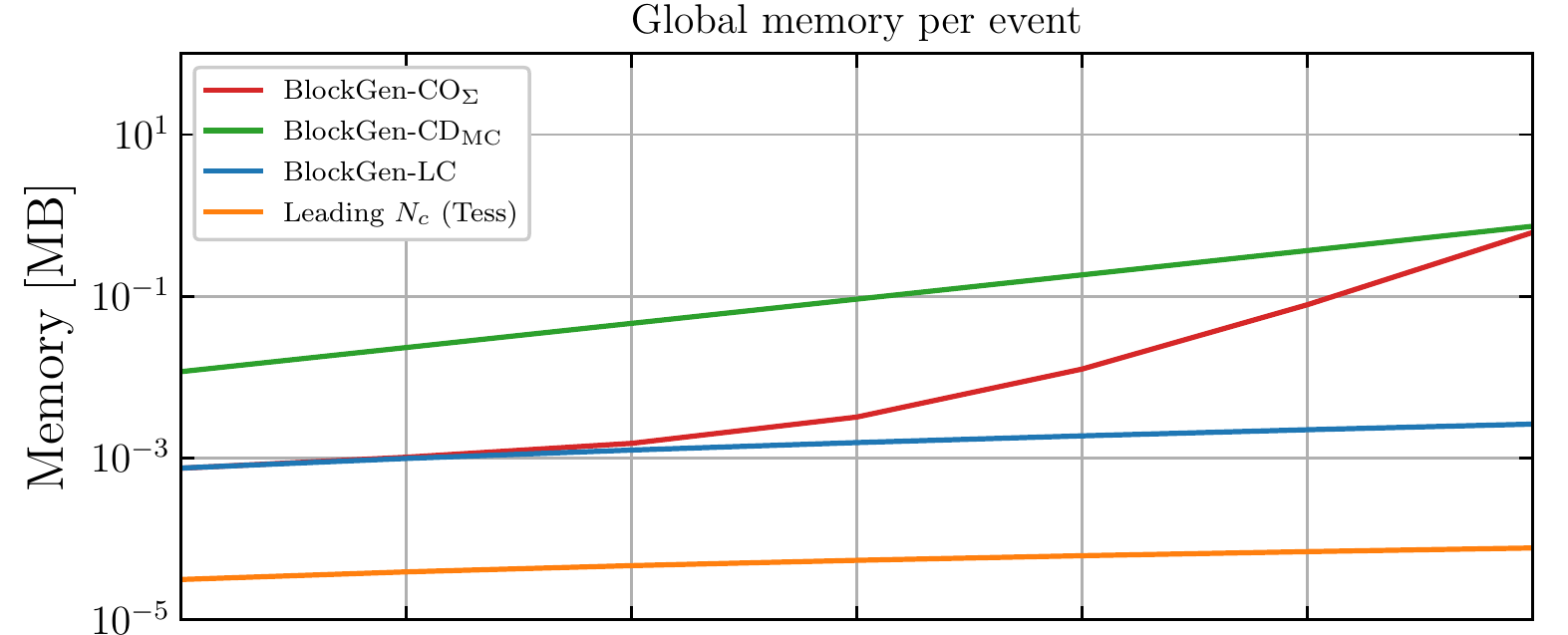}\vspace*{-.25mm}
    \includegraphics[width=\linewidth,clip,trim=0mm 0mm 0mm 132.5mm]{compare_gpu_memory.pdf}
    \end{minipage}
    \begin{minipage}[b]{0.495\textwidth}
    \includegraphics[width=\linewidth,clip,trim=0mm 111mm 0mm 0mm]{compare_gpu_memory.pdf}
    \includegraphics[width=\linewidth,clip,trim=0mm 0mm 0mm 100mm]{compare_gpu_memory.pdf}
    \end{minipage}
    \caption{Memory scaling for typical concurrent partonic calculations as a function of number of particles.
    The top right panel shows memory independent of the number of threads, while the left and bottom right panels show the additional memory per thread. Figure is reproduced from Ref.~\cite{Bothmann:2021nch}.}
    \label{fig:blockgen_mem}
\end{figure}

\begin{figure}
    \includegraphics[width=0.495\textwidth]{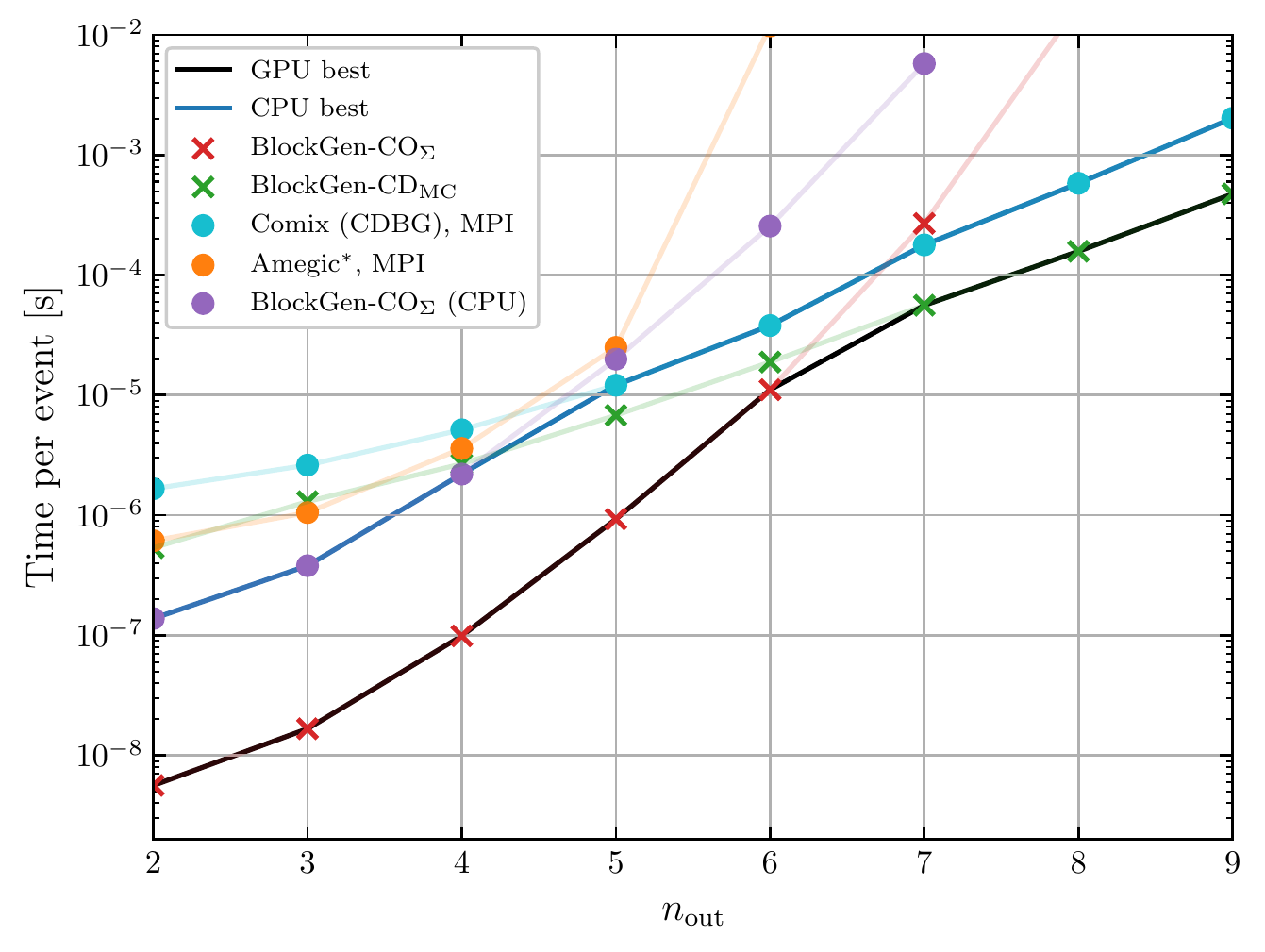}
    \begin{minipage}{0.495\textwidth}
    \vspace*{-5cm}
    \caption{The comparison of different matrix element generator implementations. The crosses denoted GPU codes, while the circles denote CPU codes. The solid black (blue) line denotes the best performance amongst all implementations on GPUs (CPUs). Figure is reproduced from Ref.~\cite{Bothmann:2021nch}.}
    \label{fig:blockgen_performance}
    \end{minipage}
\end{figure}

The \blockgen\ framework~\cite{Bothmann:2021nch} implements the Berends-Giele recursion relation~\cite{Berends:1987me,Badger:2012uz} directly onto GPUs. They performed a comprehensive study comparing color-ordered (\blockgencosummed) and color-dressed (\blockgencdsampled) implementations.
The authors found that both algorithms are memory bound and not compute bound on the GPUs. The memory required by each algorithm can be found in Fig.~\ref{fig:blockgen_mem}. The large memory footprint prevents all the needed memory fitting in the local memory, resulting in memory fetching dominating the time per event. However, significant improvement over traditional CPU implementations was found (see Fig~\ref{fig:blockgen_performance}).

\subsection{MadFlow}

\begin{figure}
    \centering
    \includegraphics[width=0.495\textwidth]{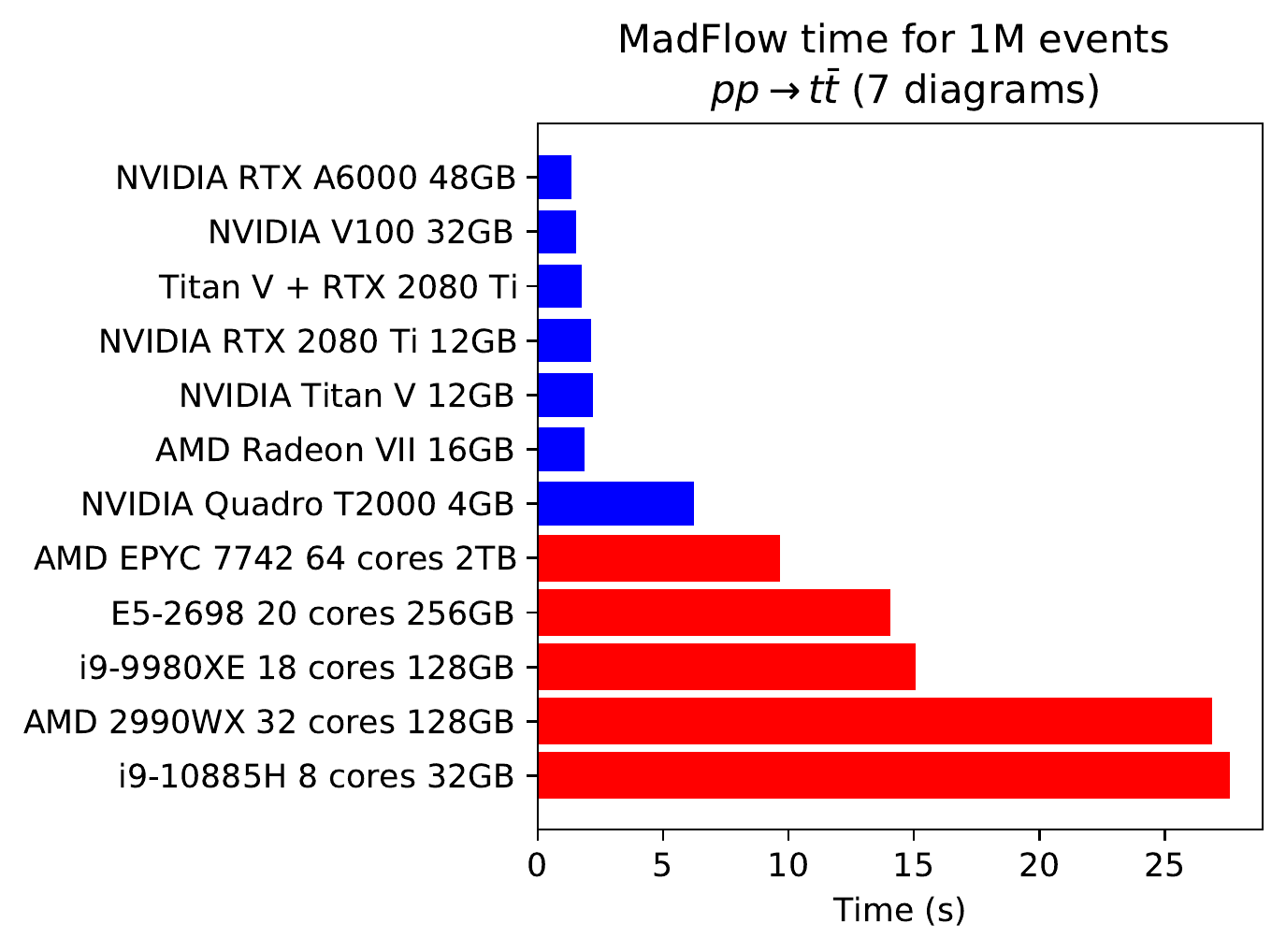}
    \includegraphics[width=0.495\textwidth]{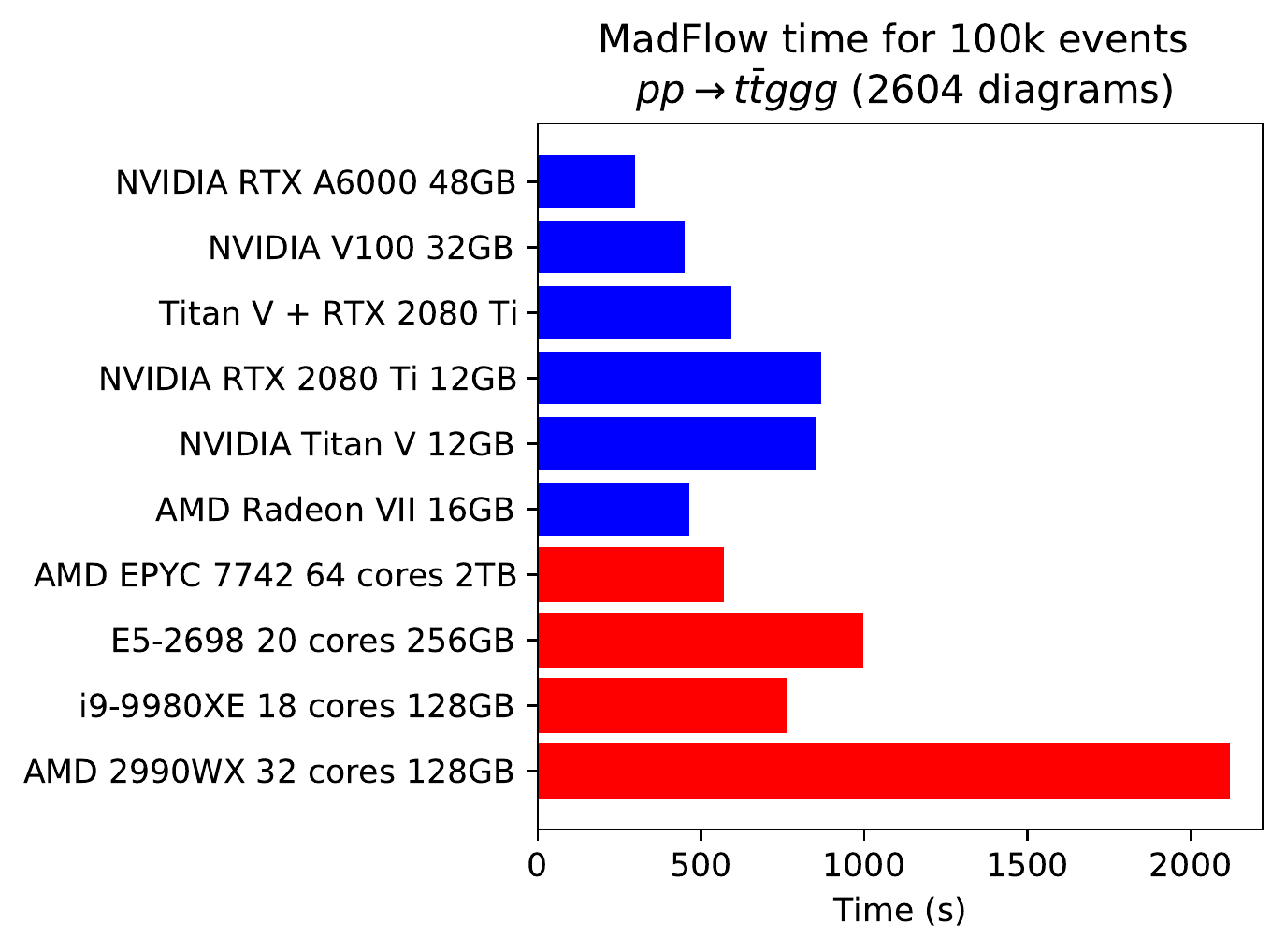}
    \caption{Performance of the MadFlow algorithm for low multiplicity (left) and high multiplicity (right). The performance of various GPUs are shown in blue, and the performance of various CPUs is shown in red. Figure is reproduced from Ref.~\cite{Carrazza:2021gpx}.}
    \label{fig:madflow}
\end{figure}

The MadFlow framework~\cite{Carrazza:2021gpx} develops an interface with the MadGraph5\_aMC@NLO program~\cite{Alwall:2014hca} to generate Tensorflow~\cite{tensorflow2015-whitepaper} code. Since Tensorflow
implements GPU calculations, the generated code from MadFlow can be directly run on either CPUs or GPUs. The MadFlow authors compared the runtime of their implementation on a variety of GPUs and CPUs to determine the overall performance improvement. While there is drastic improvements in computation speed at low multiplicities, the gains at high multiplicity are less significant (see Fig.~\ref{fig:madflow}).

\subsection{MadGraph-GPU}

The MadGraph-GPU framework builds on the MadGraph5\_aMC@NLO program to auto-generate CUDA code~\cite{Valassi:2021ljk}.
This was handled through a recursive process of optimizing the output of the C++ generator line by line, instead of redesigning the code from the ground up for GPUs. While only preliminary results are currently available for the process $e^+ e^- \rightarrow \mu^+ \mu^-$, the performance is very promising (see Tab.~\ref{tab:madgraph_gpu}).

\begin{table}[h]
    \caption{Preliminary performance results for the MadGraph-GPU code for the process $e^+e^-\rightarrow\mu^+\mu^-$. Values in the table are reproduced from Ref.~\cite{Valassi:2021ljk}.}
    \label{tab:madgraph_gpu}
\begin{center}
    \begin{tabular}{lll}
    \br
    Implementation & MEs/sec (double precision) & Ratio to C++ code \\ 
    \mr
    MadEvent Fortran & $1.50 \times 10^6$ & 1.15 \\
    Standalone C++ & $1.31 \times 10^6$ & 1.00 \\
    Standalone CUDA: NVidia V100 & $1.37 \times 10^9$ & 1050 \\
    Standalone CUDA: NVidia T4 & $4.01 \times 10^7$ & 31 \\
    \br
    \end{tabular}
\end{center}
\end{table}

\section{Machine Learning: Matrix Elements}\label{sec:ME_ML}

In addition to trying to more efficiently use the available resources on high-performance computers, other groups are investigating methods of estimating the matrix element efficiently.
These estimates can then be used to reduce the number of exact matrix evaluations required through a multiple step unweighting procedure.
This new unweighting technique is referred to as Neural Rejection Sampling~\cite{Danziger:2021eeg}.
The proposed algorithm first uses an approximation for the matrix element, and then does a first accept-reject step.
If the event fails, then there is no need to evaluate the expensive exact matrix element. However, if the event is accepted, the exact matrix element is evaluated and another accept-reject step is performed to obtain an unweighted event. The efficiency of this method can be found in Tab.~\ref{tab:neural_rejection}.

\begin{table}[h]
    \caption{Performance measures for unweighting $W+4j$ events at the LHC. $\epsilon_{\text{full}}$ is the overall efficiency, $\epsilon_{\text{(1st/2nd),surr}}$ is the efficiency for the surrogate and full matrix element respectively, and $\langle t_{\text{full}} \rangle / \langle t_{\text{surr}} \rangle$ is the ratio of time for the full matrix element to the surrogate. Further details of each parameter and data can be found in Ref.~\cite{Danziger:2021eeg}.}
    \label{tab:neural_rejection}
\begin{center}
    \begin{tabular}{lllll}
    \br
    Process & $\epsilon_{\text{full}}$ & $\langle t_{\text{full}} \rangle / \langle t_{\text{surr}} \rangle$ & $\epsilon_{\text{1st,surr}}$ & $\epsilon^{med}_{\text{2nd,surr}}$ \\
    \mr
    $dg\rightarrow e^- \bar{\nu}_e gggu$ & $1.4 \times 10^{-3}$ & 667 & $7.1 \times 10^{-4}$ & $5.3 \times 10^{-2}$ \\
    $dd\rightarrow e^- \bar{\nu}_e ggdu$ & $3.1 \times 10^{-4}$ & 162 & $1.1 \times 10^{-4}$ & $8.5 \times 10^{-2}$ \\
    $ud\rightarrow e^- \bar{\nu}_e duu\bar{d}$ & $3.6 \times 10^{-4}$ & 25 & $1.3 \times 10^{-4}$ & $7.3 \times 10^{-2}$ \\
    \br
    \end{tabular}
\end{center}
\end{table}

\begin{figure}[h]
    \centering
    \begin{minipage}[b]{0.495\textwidth}
    \includegraphics[width=\textwidth,clip,trim=0mm 100mm 120mm 0mm]{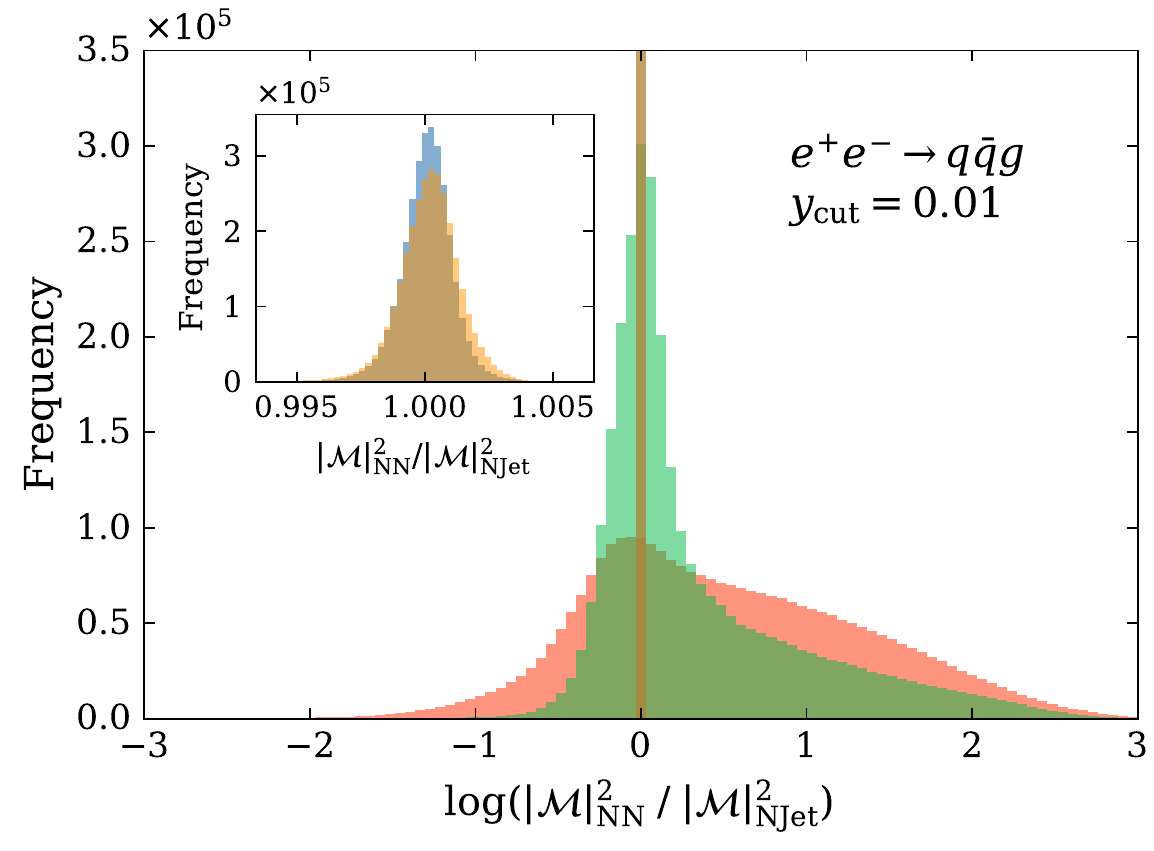}
    \end{minipage}
    \begin{minipage}[b]{0.495\textwidth}
    \includegraphics[width=\textwidth,clip,trim=120mm 12.5mm 0mm 85mm]{n3jet_comparison_inset.pdf}
    \end{minipage}
    \begin{minipage}[b]{\textwidth}
    \includegraphics[width=\textwidth,clip,trim=0mm 0mm 0mm 177mm]{n3jet_comparison_inset.pdf} 
    \end{minipage}
    \caption{Comparison of a variety of neural network estimations of the matrix element compared to the true matrix element for 3 jet (left) and 5 jet (right) production. The factorization aware result can be found in blue and orange, with a previous approach shown in red and green. Figure is reproduced from Ref.~\cite{Maitre:2021uaa}.}
    \label{fig:emulator}
\end{figure}

To obtain optimal performance of the neural rejection algortihm requires an extremely accurate matrix element estimate. This can be obtained with the help of a factorization aware Matrix element emulator. Such an emulator has been proposed in Ref.~\cite{Maitre:2021uaa}. It uses information about the physics to develop an extremely accurate neural network estimate. The estimate takes advantage of the known factorization properties of the the matrix element in the soft and collinear limits~\cite{Catani:1996vz}. Using this information, an ansatz can be fit through the use of a neural network. The accuracy of this method compared to previous results can be found in Fig.~\ref{fig:emulator}.

\section{Machine Learning: Event Generation}\label{sec:PS_ML}

While improving the computational speed of the unweighting procedure can help to drastically reduce the computational cost, another avenue is in using machine learning to generate events.
There are two approaches in this direction: generating events from a set of pre-generated events~\cite{Bendavid:2017zhk,Otten:2019hhl,Hashemi:2019fkn,DiSipio:2019imz,Butter:2019cae,Carrazza:2019cnt,SHiP:2019gcl,Butter:2019eyo,Butter:2020qhk,Butter:2020tvl}, and generating the phase space efficiently~\cite{Bendavid:2017zhk,Klimek:2018mza,Bothmann:2020ywa,Gao:2020vdv,Gao:2020zvv,Chen:2020nfb,Stienen:2020gns}.

\subsection{From Events}

\begin{figure}
    \includegraphics[width=0.3\textwidth]{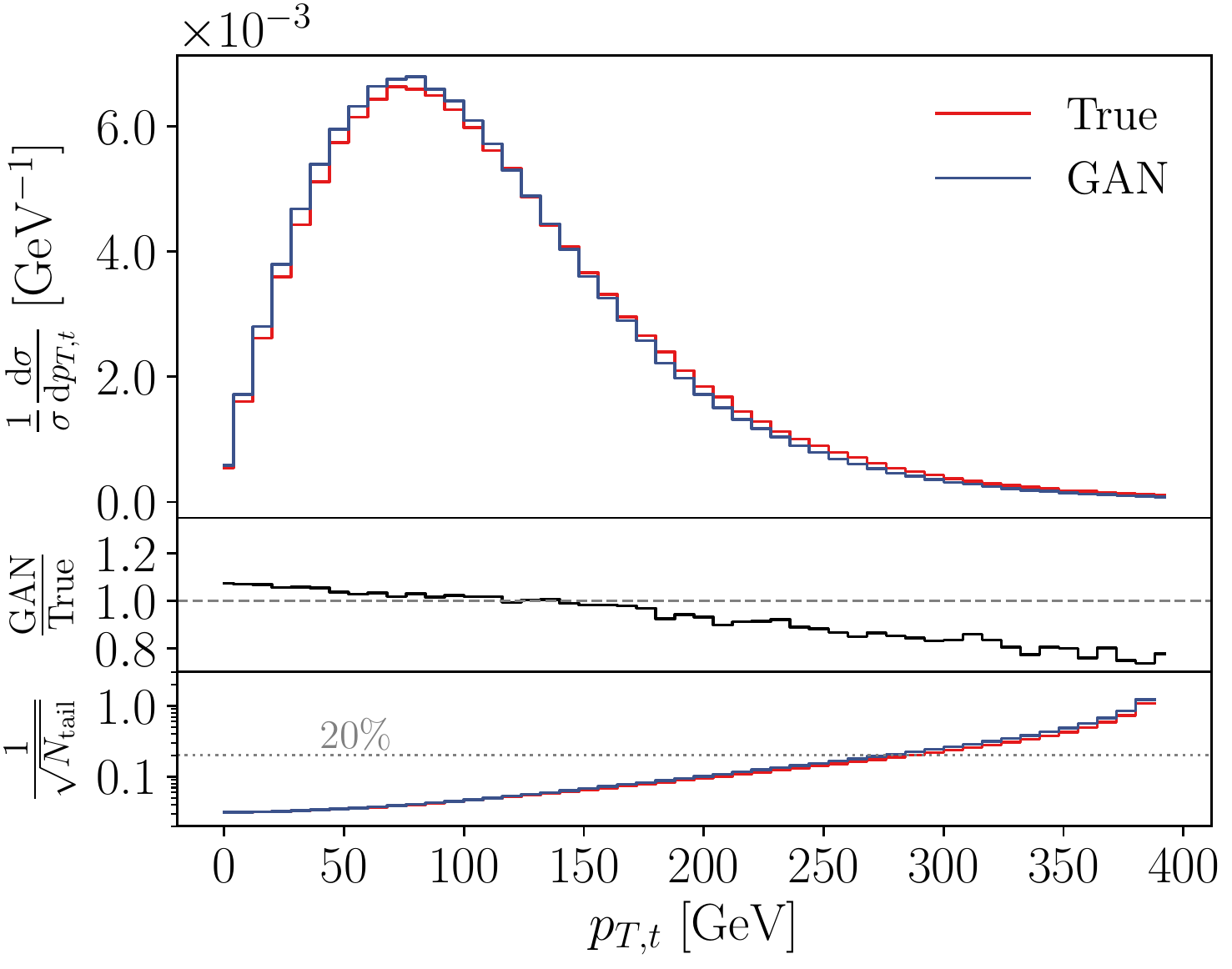}
    \hfill
    \includegraphics[width=0.3\textwidth]{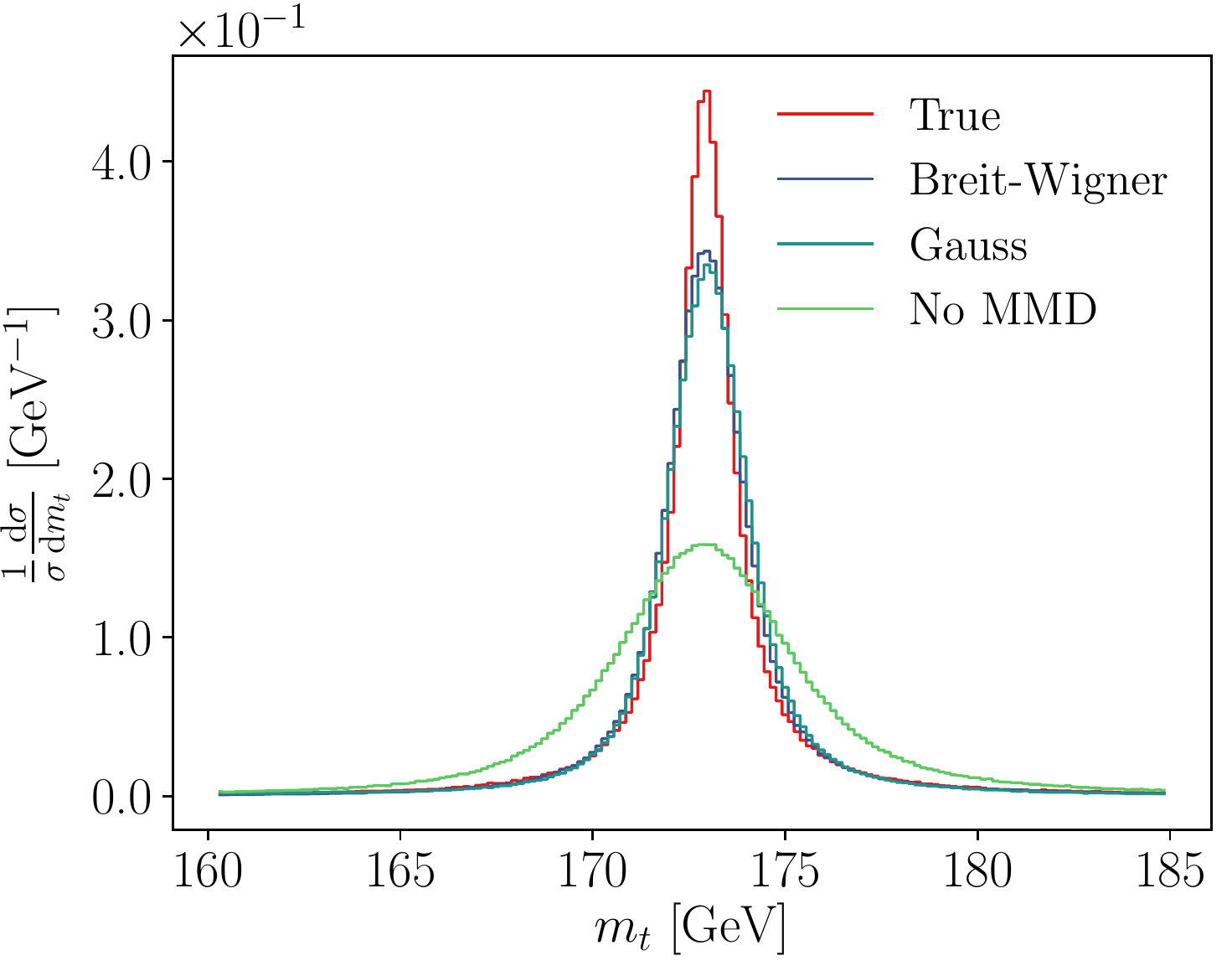}
    \hfill
    \includegraphics[width=0.3\textwidth]{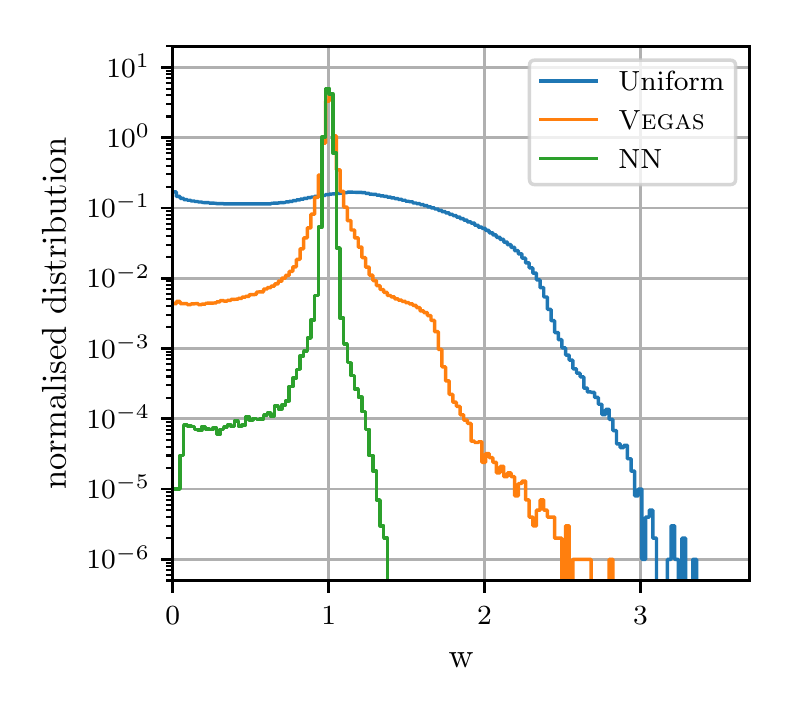}
    \caption{Left and Center: Event generation through the use of a trained GAN network. The left plot shows the accuracy of GANs for smooth distributions, while the center plot shows the accuracy of these techniques for peaked distributions. The different curves in the center plot are modifications added to the loss to improve the accuracy.
    Right: The weight distribution for even generation. The unweighting efficiency is directly related to the narrowness of the distribution. The neural network (green) shows a drastic improvement over the previous optimal method VEGAS (orange).
    Left and center figures are reproduced from Ref.~\cite{Butter:2019cae}, and the right figure is reproduced from Ref.~\cite{Bothmann:2020ywa}.}
    \label{fig:mc_gan}
\end{figure}

The generation of events from a set of pre-generated events are able to quickly generate additional events that are similar to the existing ones.
However, these approaches have no guarantees on maintaining the total cross section and struggle with peaked structures in the integrand. The work in this direction mainly focuses on the use of GANs to generate the events, and is very successful in reproducing smooth distributions. The current state of the art for these methods can be seen in Fig.~\ref{fig:mc_gan}.

\subsection{Phase Space}

The other approach attempts to use machine learning to greatly increase the unweighting efficiency. The advantage of this approach is that it does not require a pre-generated sample and through the unweighting procedure guarantees to reproduce the correct cross section. The downside of this approach is that the networks are more computationally intensive, and still requires an unweighting step. These two issues lead to a drastically slower method of generating events than the above approach. The main architecture investigated are invertible networks, and some promising results have been found in Refs.~\cite{Bothmann:2020ywa,Gao:2020vdv,Gao:2020zvv} (see the right panel of Fig.~\ref{fig:mc_gan} and Tab.~\ref{tab:iflow}).

\begin{table}[h]
    \caption{Unweighting efficiencies for the default Sherpa and the improvements from neural network phase space generation for a variety of processes of interest. Data for the table is taken from Ref.~\cite{Gao:2020zvv}.}
    \label{tab:iflow}
\begin{center}
    \begin{tabular}{lllllll}
    \br
    \multicolumn{2}{l}{unweighting efficiency} & \multicolumn{5}{l}{LO QCD} \\
    & & $n=0$ & $n=1$ & $n=2$ & $n=3$ & $n=4$ \\
    \mr
    $W^+ + n$ jets & Sherpa & $2.8\times 10^{-1}$ & $3.8\times 10^{-2}$ & $7.5 \times 10^{-3}$ & $1.5\times 10^{-3}$ & $8.3 \times 10^{-4}$ \\
    & NN+NF & $6.1 \times 10^{-1}$ & $1.2 \times 10^{-1}$ & $1.0 \times 10^{-2}$ & $1.8 \times 10^{-3}$ & $8.9 \times 10^{-4}$ \\
    & Gain & 2.2 & 3.3 & 1.4 & 1.2 & 1.1 \\
    \mr
    $W^- + n$ jets & Sherpa & $2.9\times 10^{-1}$ & $4.0\times 10^{-2}$ & $7.7 \times 10^{-3}$ & $2.0\times 10^{-3}$ & $9.7 \times 10^{-4}$ \\
    & NN+NF & $7.0 \times 10^{-1}$ & $1.5 \times 10^{-1}$ & $1.1 \times 10^{-2}$ & $2.2 \times 10^{-3}$ & $7.9 \times 10^{-4}$ \\
    & Gain & 2.4 & 3.3 & 1.4 & 1.1 & 0.82 \\
    \mr
    $Z^- + n$ jets & Sherpa & $3.1\times 10^{-1}$ & $3.6\times 10^{-2}$ & $1.5 \times 10^{-2}$ & $4.7\times 10^{-3}$ &  \\
    & NN+NF & $3.8 \times 10^{-1}$ & $1.0 \times 10^{-1}$ & $1.4 \times 10^{-2}$ & $2.4 \times 10^{-3}$ &  \\
    & Gain & 1.2 & 2.9 & 0.91 & 0.51 &  \\
    \br
    \end{tabular}
\end{center}
\end{table}

\section{Reducing Negative Weights}\label{sec:negative}

In higher order calculations, negative event weights are unavoidable. 
Negative weights arise in the color dipole terms, overestimation of real-emission matrix elements, and in matching to parton showers.
The impact of negative weight events is a reduction in effective sample size, and thus statistical accuracy. Given a set of events with weights either positive one or negative one, with a fraction of $\epsilon$ events with negative weight, the additional events required for the same accuracy is given as:
\begin{equation}
    f(\epsilon) = \frac{1}{(1-2\epsilon)^2}.
\end{equation}
This quickly increases as $\epsilon$ approaches 50\%. Therefore, to improve the efficiency of event generation, the fraction of negative weight events needs to be as small as possible. There are two methods of reducing the negative weight fraction.

The first is through the use of resampling the events to produce only positive weight events. This can be accomplished through the use of a positiver resampler~\cite{Andersen:2020sjs}, a neural network resampler~\cite{Nachman:2020fff}, or cell resampling~\cite{Andersen:2021mvw}.

In the positiver resampler~\cite{Andersen:2020sjs}, events are rescaled by $\frac{\sum_i w_i}{\sum_i |w_i|}$, partially unweighted, and then rescaled to ensure that the original histogram bin height is restored. The reduction in events required to reach a given precision can be seen in Tab.~\ref{tab:resample}.

\begin{table}[h]
    \caption{Number of weighted events generated and number of events that pass analysis cuts from merged $W+0j,1j,2j$ at next-to-leading order calculation. Data is taken from Ref.~\cite{Andersen:2020sjs}.}
    \label{tab:resample}
\begin{center}
    \begin{tabular}{lll}
    \br
    Sample & Total number of events & Events included in analysis \\
    \mr
    weighted & 5.3M & 195k \\
    positive only & 3.2M & 121k \\
    unweighted & 1.5M & 52k \\
    Positive Resampler($t$) & 659k & 25k \\
    Positive Resampler($p_T^W$) & 33k & 33k \\
    \br
    \end{tabular}
\end{center}
\end{table}

\begin{figure}[h]
    \includegraphics[width=0.4\textwidth]{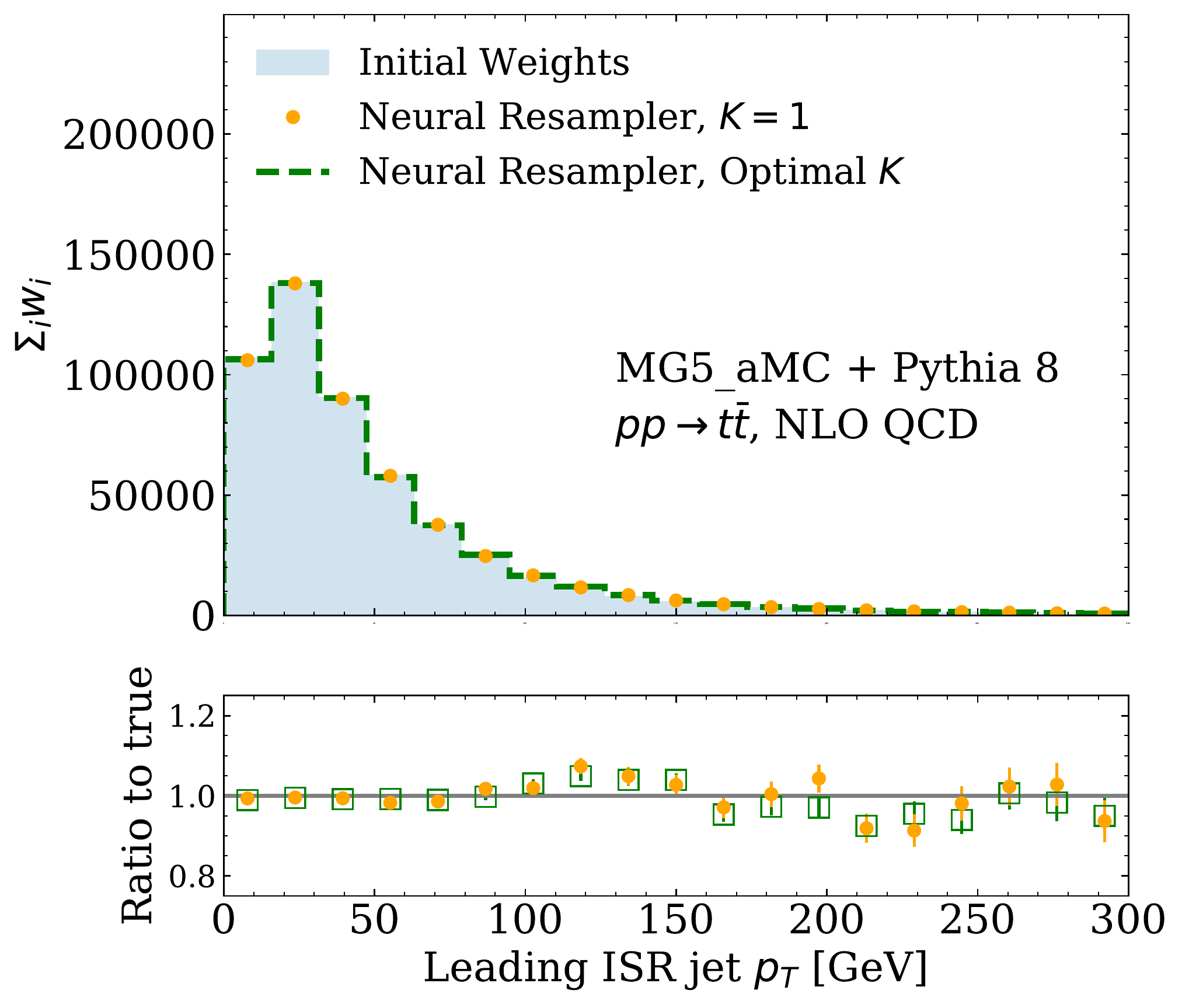}
    \hfill
    \includegraphics[width=0.495\textwidth]{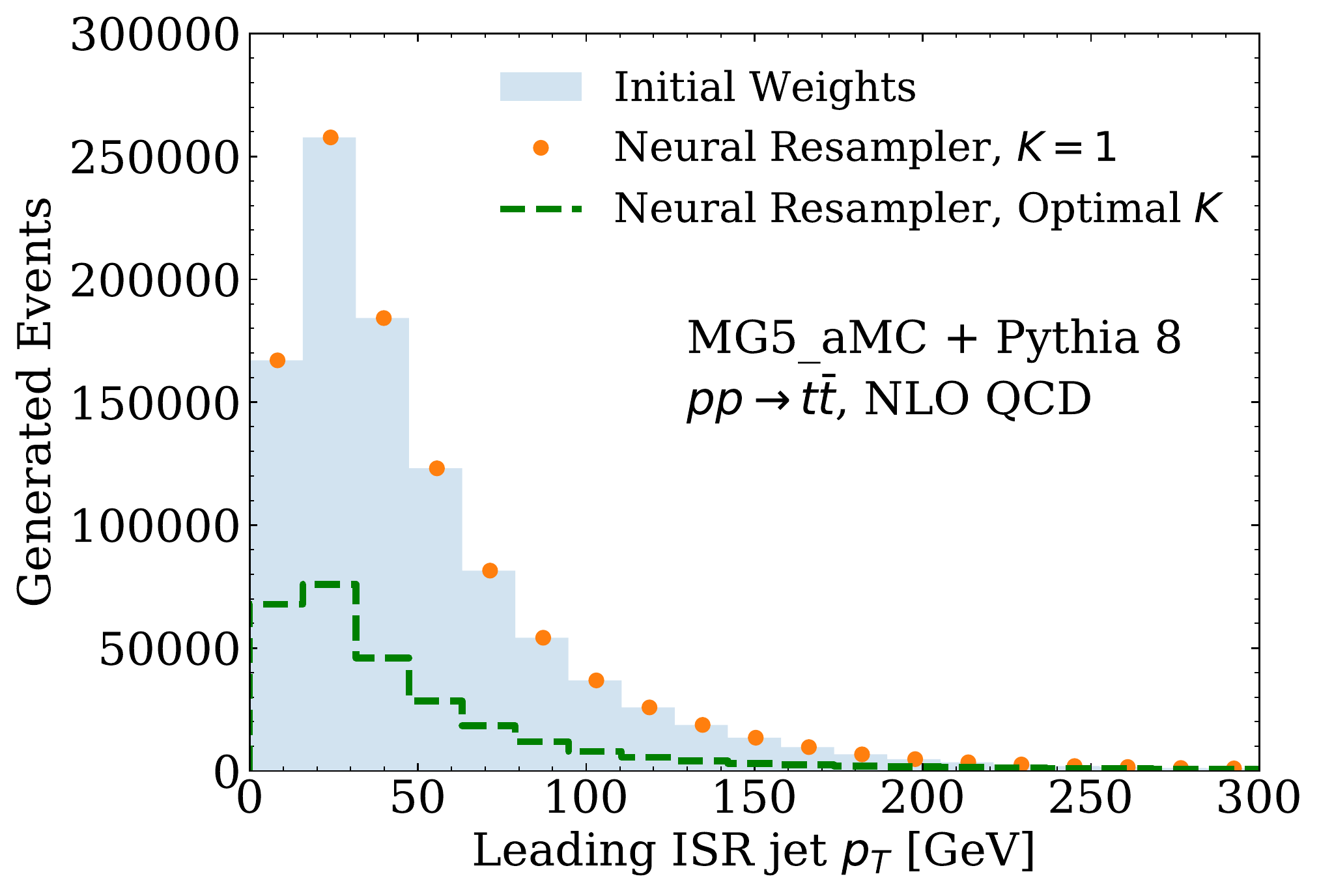}
    \caption{Reduction in required number of events generated through a neural resampling procedure. Figure is reproduced from Ref.~\cite{Nachman:2020fff}.}
    \label{fig:resample}
\end{figure}

In the neural resampler~\cite{Nachman:2020fff}, a weight and variance are obtained through a neural network to estimate how close the result is to uniformly weighted sample. The weight and variance are then used to calculate a scaling factor. This scaling factor is used to perform an accept-reject step and then the event weight is increased by this scaling factor. One advantage to this technique is that the variance of the original sample is maintained. The validation and reduction in events generated for $pp\rightarrow t\bar{t}$ at NLO is shown in Fig.~\ref{fig:resample}.

\begin{figure}
    \includegraphics[width=0.4\textwidth]{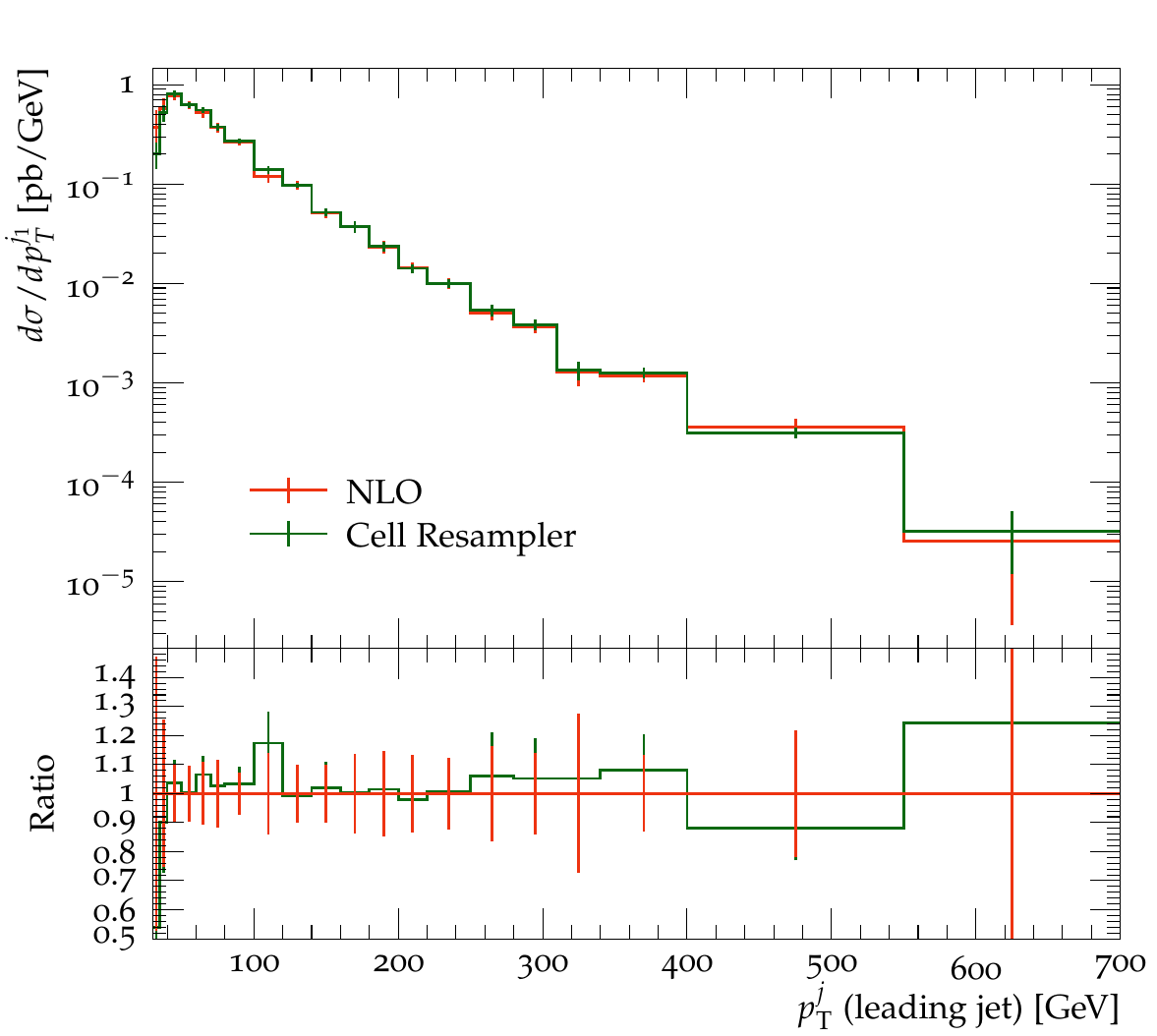}
    \hfill
    \includegraphics[width=0.495\textwidth]{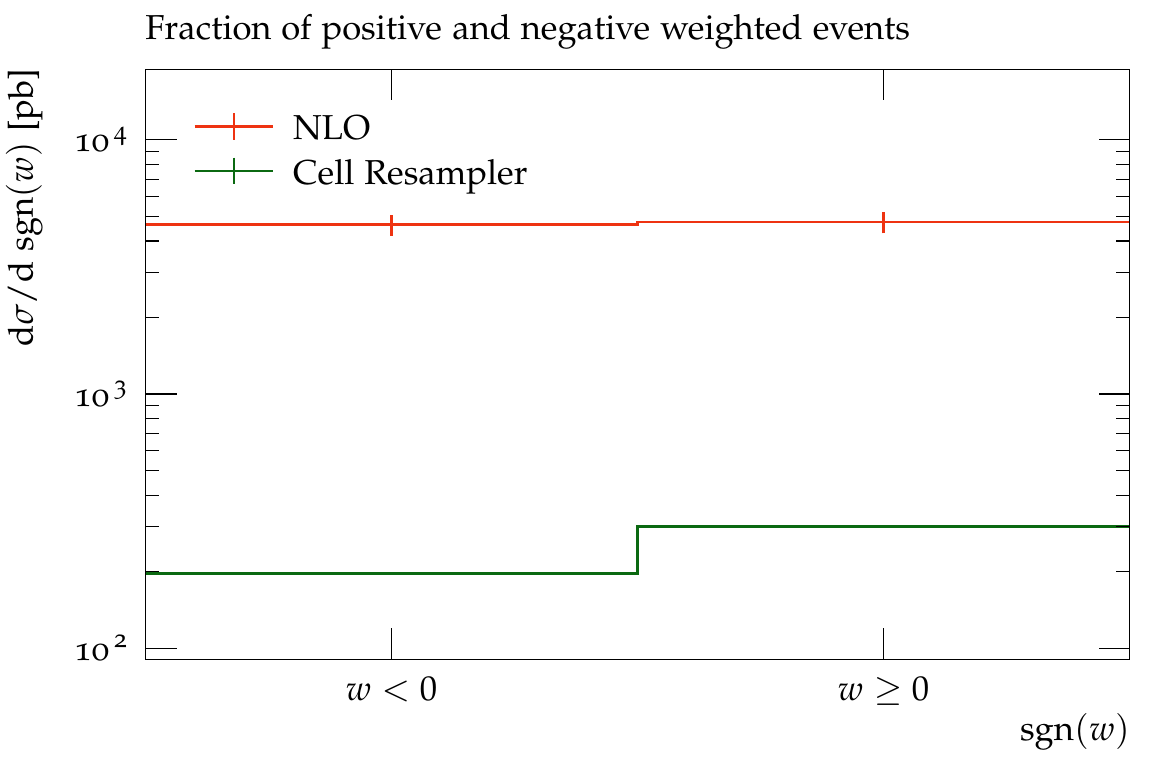}
    \caption{Left: Validation that the cell resampler reproduces the expected result. Right: The fraction of positive weight and negative weight events before and after the cell resampler. The figure is reproduced from Ref.~\cite{Andersen:2021mvw}.}
    \label{fig:cell_resample}
\end{figure}

In the cell resampler~\cite{Andersen:2021mvw}, a phase space measure is defined to calculate the distance between two points. The negative weight events are then chosen as a seed of a cell. The cell size is increased until the sum of the weights in the cell is at least twice the negative weight seed. Afterwards, all the events in the cell are scaled $\frac{\sum_{i\in\mathcal{C}} w_i}{\sum_{i\in\mathcal{C}} |w_i|}$. If there is an insufficient number of events in a local region to make the weights positive, the negative weight is kept. This drastically decreases the number of events required and the fraction of negative weights (see Fig.~\ref{fig:cell_resample}).

Finally, in addition to resampling techniques, the algorithms can be improved to reduce the negative weight fraction. In Ref.~\cite{Danziger:2021xvr}, three improvements were proposed. These improvements were matching to leading color showers, shower vetos on hard emission events, and local K-factors from the core process. The combination of these three improvements halved the negative weight fraction within the study. The breakdown of the effect of each improvement can be found in Tab.~\ref{tab:sherpa_neg}.

\begin{table}[h]
    \caption{Improvements made in the Sherpa event generator to reduce the fraction of negative weight events. Data is taken from Ref.~\cite{Danziger:2021xvr}.}
    \label{tab:sherpa_neg}
    \centering
    \begin{tabular}{ll}
    \br
     & Negative Weight Fraction \\
    \mr
    Default & 18.1\% \\
    \mr
    Leading Color Mode & 14.0\% \\
    + shower veto on $\mathbb{H}$-events & 9.6\% \\
    + local K-factor from core & 9.1\% \\
    \br
    \end{tabular}
\end{table}

\section{Conclusions}

In order to match the statistical precision of the high-luminosity LHC, a large number of events need to be generated with the help of MC programs. The currently available event generators require a computing budget that exceeds the provided budget to reach the desired goal. Therefore, there have been many steps taken to improve event generators to reduce the cost per event.

Speed improvements through the use of dedicated GPU implementations for the hard matrix element show promise in reducing the cost, but may not be sufficient on their own. Using machine learning techniques to provide faster unweighting methods and improved phase space generation also help to increase the overall speed of event generation. Finally, studying negative weight reduction techniques works on minimizing number of simulated events required.

While each of these steps help to work towards reducing the computing costs to the desired level, there is still more that needs to be done in order to meet the experimental requirements.

\ack

I would like to thank Stefan H{\"o}che for useful discussions and advice on the preparation of these proceedings.
This manuscript has been authored by Fermi Research Alliance, LLC under Contract No. DE-AC02-07CH11359 with the U.S. Department of Energy, Office of Science, Office of High Energy Physics.

\bibliography{biblio}

\providecommand{\newblock}{}
\begin{thebibliography}{10}
\expandafter\ifx\csname url\endcsname\relax
  \def\url#1{{\tt #1}}\fi
\expandafter\ifx\csname urlprefix\endcsname\relax\def\urlprefix{URL }\fi
\providecommand{\eprint}[2][]{\url{#2}}

\bibitem{Webber:1986mc}
Webber B 1986 {\em Ann. Rev. Nucl. Part. Sci.\/} {\bf 36} 253--286
  \urlprefix\url{http://inspirebeta.net/record/18125}

\bibitem{Buckley:2011ms}
Buckley A {\em et~al.\/} 2011 {\em Phys. Rept.\/} {\bf 504} 145--233
  (\textit{Preprint} \eprint{1101.2599})
  \urlprefix\url{http://inspirebeta.net/record/884202}

\bibitem{Kanaki:2000ey}
Kanaki A and Papadopoulos C~G 2000 {\em Comput. Phys. Commun.\/} {\bf 132}
  306--315 (\textit{Preprint} \eprint{hep-ph/0002082})
  \urlprefix\url{http://inspirehep.net/search?p=hep-ph/0002082}

\bibitem{Papadopoulos:2000tt}
Papadopoulos C~G 2001 {\em Comput. Phys. Commun.\/} {\bf 137} 247--254
  (\textit{Preprint} \eprint{hep-ph/0007335})
  \urlprefix\url{http://inspirehep.net/search?p=hep-ph/0007335}

\bibitem{Krauss:2001iv}
Krauss F, Kuhn R and Soff G 2002 {\em JHEP\/} {\bf 02} 044 (\textit{Preprint}
  \eprint{hep-ph/0109036})
  \urlprefix\url{http://inspirehep.net/search?p=hep-ph/0109036}

\bibitem{Gleisberg:2008fv}
Gleisberg T and H{\"o}che S 2008 {\em JHEP\/} {\bf 12} 039 (\textit{Preprint}
  \eprint{0808.3674}) \urlprefix\url{http://inspirehep.net/record/793879}

\bibitem{Maltoni:2002qb}
Maltoni F and Stelzer T 2003 {\em JHEP\/} {\bf 02} 027 (\textit{Preprint}
  \eprint{hep-ph/0208156})
  \urlprefix\url{http://inspirehep.net/search?p=hep-ph/0208156}

\bibitem{Alwall:2007st}
Alwall J {\em et~al.\/} 2007 {\em JHEP\/} {\bf 09} 028 (\textit{Preprint}
  \eprint{0706.2334})
  \urlprefix\url{http://inspirehep.net/search?p=arXiv:0706.2334}

\bibitem{Alwall:2011uj}
Alwall J, Herquet M, Maltoni F, Mattelaer O and Stelzer T 2011 {\em JHEP\/}
  {\bf 06} 128 (\textit{Preprint} \eprint{1106.0522})
  \urlprefix\url{http://inspirebeta.net/record/912611}

\bibitem{Mangano:2002ea}
Mangano M~L, Moretti M, Piccinini F, Pittau R and Polosa A~D 2003 {\em JHEP\/}
  {\bf 07} 001 (\textit{Preprint} \eprint{hep-ph/0206293})
  \urlprefix\url{http://inspirehep.net/search?p=hep-ph/0206293}

\bibitem{Ossola:2007ax}
Ossola G, Papadopoulos C~G and Pittau R 2008 {\em JHEP\/} {\bf 0803} 042
  (\textit{Preprint} \eprint{0711.3596})

\bibitem{Gleisberg:2007md}
Gleisberg T and Krauss F 2008 {\em Eur. Phys. J.\/} {\bf C53} 501--523
  (\textit{Preprint} \eprint{0709.2881})
  \urlprefix\url{http://inspirehep.net/search?p=arXiv:0709.2881}

\bibitem{Berger:2008sj}
Berger C~F, Bern Z, Dixon L~J, Febres-Cordero F, Forde D, Ita H, Kosower D~A
  and Ma{\^i}tre D 2008 {\em Phys. Rev.\/} {\bf D78} 036003 (\textit{Preprint}
  \eprint{0803.4180}) \urlprefix\url{http://inspirebeta.net/record/782271}

\bibitem{Bevilacqua:2011xh}
Bevilacqua G, Czakon M, Garzelli M, van Hameren A, Kardos A {\em et~al.\/} 2013
  {\em Comput.Phys.Commun.\/} {\bf 184} 986--997 (\textit{Preprint}
  \eprint{1110.1499})

\bibitem{Cascioli:2011va}
Cascioli F, Maierh{\"o}fer P and Pozzorini S 2012 {\em Phys.Rev.Lett.\/} {\bf
  108} 111601 (\textit{Preprint} \eprint{1111.5206})
  \urlprefix\url{http://inspirehep.net/record/946998}

\bibitem{Cullen:2014yla}
Cullen G, van Deurzen H, Greiner N, Heinrich G, Luisoni G {\em et~al.\/} 2014
  {\em Eur.Phys.J.\/} {\bf C74} 3001 (\textit{Preprint} \eprint{1404.7096})
  \urlprefix\url{http://inspirehep.net/record/1292822}

\bibitem{Alwall:2014hca}
Alwall J, Frederix R, Frixione S, Hirschi V, Maltoni F, Mattelaer O, Shao H~S,
  Stelzer T, Torrielli P and Zaro M 2014 {\em JHEP\/} {\bf 07} 079
  (\textit{Preprint} \eprint{1405.0301})

\bibitem{Actis:2016mpe}
Actis S, Denner A, Hofer L, Lang J~N, Scharf A and Uccirati S 2017 {\em Comput.
  Phys. Commun.\/} {\bf 214} 140--173 (\textit{Preprint} \eprint{1605.01090})

\bibitem{Christensen:2009jx}
Christensen N~D, de~Aquino P, Degrande C, Duhr C, Fuks B, Herquet M, Maltoni F
  and Schumann S 2011 {\em Eur. Phys. J.\/} {\bf C71} 1541 (\textit{Preprint}
  \eprint{0906.2474}) \urlprefix\url{http://inspirebeta.net/record/823106}

\bibitem{Degrande:2011ua}
Degrande C, Duhr C, Fuks B, Grellscheid D, Mattelaer O and Reiter T 2012 {\em
  Comput.Phys.Commun.\/} {\bf 183} 1201--1214 (\textit{Preprint}
  \eprint{1108.2040})

\bibitem{Staub:2013tta}
Staub F 2014 {\em Comput.Phys.Commun.\/} {\bf 185} 1773--1790
  (\textit{Preprint} \eprint{1309.7223})

\bibitem{Calafiura:2729668}
Calafiura P, Catmore J, Costanzo D and Di~Girolamo A 2020 {ATLAS HL-LHC
  Computing Conceptual Design Report} Tech. Rep. CERN-LHCC-2020-015. LHCC-G-178
  CERN Geneva \urlprefix\url{https://cds.cern.ch/record/2729668}

\bibitem{Hoche:2019flt}
H\"oche S, Prestel S and Schulz H 2019 {\em Phys. Rev. D\/} {\bf 100} 014024
  (\textit{Preprint} \eprint{1905.05120})

\bibitem{Bothmann:2021nch}
Bothmann E, Giele W, Hoeche S, Isaacson J and Knobbe M 2021  (\textit{Preprint}
  \eprint{2106.06507})

\bibitem{Carrazza:2021gpx}
Carrazza S, Cruz-Martinez J, Rossi M and Zaro M 2021 {\em Eur. Phys. J. C\/}
  {\bf 81} 656 (\textit{Preprint} \eprint{2106.10279})

\bibitem{Valassi:2021ljk}
Valassi A, Roiser S, Mattelaer O and Hageboeck S 2021 {\em EPJ Web Conf.\/}
  {\bf 251} 03045 (\textit{Preprint} \eprint{2106.12631})

\bibitem{Berends:1987me}
Berends F~A and Giele W~T 1988 {\em Nucl. Phys.\/} {\bf B306} 759
  \urlprefix\url{http://inspirehep.net/search?j=NUPHA,B306,759}

\bibitem{Badger:2012uz}
Badger S, Biedermann B, Hackl L, Plefka J, Schuster T and Uwer P 2013 {\em
  Phys. Rev.\/} {\bf D87} 034011 (\textit{Preprint} \eprint{1206.2381})

\bibitem{tensorflow2015-whitepaper}
Abadi M, Agarwal A, Barham P, Brevdo E, Chen Z, Citro C, Corrado G~S, Davis A,
  Dean J, Devin M, Ghemawat S, Goodfellow I, Harp A, Irving G, Isard M, Jia Y,
  Jozefowicz R, Kaiser L, Kudlur M, Levenberg J, Man\'{e} D, Monga R, Moore S,
  Murray D, Olah C, Schuster M, Shlens J, Steiner B, Sutskever I, Talwar K,
  Tucker P, Vanhoucke V, Vasudevan V, Vi\'{e}gas F, Vinyals O, Warden P,
  Wattenberg M, Wicke M, Yu Y and Zheng X 2015 {TensorFlow}: Large-scale
  machine learning on heterogeneous systems software available from
  tensorflow.org \urlprefix\url{https://www.tensorflow.org/}

\bibitem{Danziger:2021eeg}
Danziger K, Jan\ss{}en T, Schumann S and Siegert F 2021  (\textit{Preprint}
  \eprint{2109.11964})

\bibitem{Maitre:2021uaa}
Ma\^\i{}tre D and Truong H 2021 {\em JHEP\/} {\bf 11} 066 (\textit{Preprint}
  \eprint{2107.06625})

\bibitem{Catani:1996vz}
Catani S and Seymour M~H 1997 {\em Nucl. Phys.\/} {\bf B485} 291--419
  (\textit{Preprint} \eprint{hep-ph/9605323})
  \urlprefix\url{http://inspirehep.net/search?p=hep-ph/9605323}

\bibitem{Bendavid:2017zhk}
Bendavid J 2017  (\textit{Preprint} \eprint{1707.00028})

\bibitem{Otten:2019hhl}
Otten S, Caron S, de~Swart W, van Beekveld M, Hendriks L, van Leeuwen C,
  Podareanu D, Ruiz~de Austri R and Verheyen R 2021 {\em Nature Commun.\/} {\bf
  12} 2985 (\textit{Preprint} \eprint{1901.00875})

\bibitem{Hashemi:2019fkn}
Hashemi B, Amin N, Datta K, Olivito D and Pierini M 2019  (\textit{Preprint}
  \eprint{1901.05282})

\bibitem{DiSipio:2019imz}
Di~Sipio R, Faucci~Giannelli M, Ketabchi~Haghighat S and Palazzo S 2019 {\em
  JHEP\/} {\bf 08} 110 (\textit{Preprint} \eprint{1903.02433})

\bibitem{Butter:2019cae}
Butter A, Plehn T and Winterhalder R 2019 {\em SciPost Phys.\/} {\bf 7} 075
  (\textit{Preprint} \eprint{1907.03764})

\bibitem{Carrazza:2019cnt}
Carrazza S and Dreyer F~A 2019 {\em Eur. Phys. J. C\/} {\bf 79} 979
  (\textit{Preprint} \eprint{1909.01359})

\bibitem{SHiP:2019gcl}
Ahdida C {\em et~al.\/} (SHiP) 2019 {\em JINST\/} {\bf 14} P11028
  (\textit{Preprint} \eprint{1909.04451})

\bibitem{Butter:2019eyo}
Butter A, Plehn T and Winterhalder R 2019  (\textit{Preprint}
  \eprint{1912.08824})

\bibitem{Butter:2020qhk}
Butter A, Diefenbacher S, Kasieczka G, Nachman B and Plehn T 2021 {\em SciPost
  Phys.\/} {\bf 10} 139 (\textit{Preprint} \eprint{2008.06545})

\bibitem{Butter:2020tvl}
Butter A and Plehn T 2020  (\textit{Preprint} \eprint{2008.08558})

\bibitem{Klimek:2018mza}
Klimek M~D and Perelstein M 2020 {\em SciPost Phys.\/} {\bf 9} 053
  (\textit{Preprint} \eprint{1810.11509})

\bibitem{Bothmann:2020ywa}
Bothmann E, Jan\ss{}en T, Knobbe M, Schmale T and Schumann S 2020 {\em SciPost
  Phys.\/} {\bf 8} 069 (\textit{Preprint} \eprint{2001.05478})

\bibitem{Gao:2020vdv}
Gao C, Isaacson J and Krause C 2020 {\em Mach. Learn. Sci. Tech.\/} {\bf 1}
  045023 (\textit{Preprint} \eprint{2001.05486})

\bibitem{Gao:2020zvv}
Gao C, H\"oche S, Isaacson J, Krause C and Schulz H 2020 {\em Phys. Rev. D\/}
  {\bf 101} 076002 (\textit{Preprint} \eprint{2001.10028})

\bibitem{Chen:2020nfb}
Chen I~K, Klimek M~D and Perelstein M 2021 {\em SciPost Phys.\/} {\bf 10} 023
  (\textit{Preprint} \eprint{2009.07819})

\bibitem{Stienen:2020gns}
Stienen B and Verheyen R 2021 {\em SciPost Phys.\/} {\bf 10} 038
  (\textit{Preprint} \eprint{2011.13445})

\bibitem{Andersen:2020sjs}
Andersen J~R, G\"utschow C, Maier A and Prestel S 2020 {\em Eur. Phys. J. C\/}
  {\bf 80} 1007 (\textit{Preprint} \eprint{2005.09375})

\bibitem{Nachman:2020fff}
Nachman B and Thaler J 2020 {\em Phys. Rev. D\/} {\bf 102} 076004
  (\textit{Preprint} \eprint{2007.11586})

\bibitem{Andersen:2021mvw}
Andersen J~R and Maier A 2021  (\textit{Preprint} \eprint{2109.07851})

\bibitem{Danziger:2021xvr}
Danziger K, H\"oche S and Siegert F 2021  (\textit{Preprint}
  \eprint{2110.15211})

\end{thebibliography}

\end{document}